\documentclass[11pt]{article}
\title{Encoding of (part of) N=4 superconformal algebra into 4 twisted differential operators}
\usepackage{mathrsfs}
\usepackage{amsmath}

\global\arraycolsep=1pt
\usepackage[colorlinks=true,backref=true,linkcolor=black,anchorcolor=black,citecolor=black,filecolor=black,menucolor=black,pagecolor=black,urlcolor=black]{hyperref}
\usepackage[Symbol]{upgreek}
\usepackage{bbm}
\usepackage{dsfont}
\usepackage{amssymb}
\usepackage{textcomp}

\newcommand{\scal}[1]{\bigl ({#1} \bigr )}
\def\bea{\begin{eqnarray}}
\def\eea{\end{eqnarray}}
\def\be{\begin{equation}}
\def\ee{\end{equation}}
\newcommand{\CR}{\nonumber \\*}

\newcommand{\trace}{\hbox {Tr}~}

\def\L{{\cal L}}

\def\Lc{\mathscr{L}}

\def\a{\mathfrak{a}}
\def\b{\mathfrak{b}}

\def\d{\mathfrak{d}}
\def\e{\mathfrak{e}}
\def\f{\mathfrak{f}}

\usepackage[colorlinks=true,backref=true,linkcolor=black,anchorcolor=black,citecolor=black,filecolor=black,menucolor=black,pagecolor=black,urlcolor=black]{hyperref}

\usepackage[Symbol]{upgreek}
\usepackage{caption}
\usepackage{bbm}
\usepackage{dsfont}
\usepackage{amssymb}
\usepackage{textcomp}
\def\bea{\begin{eqnarray}}
\def\eea{\end{eqnarray}}
\def\be{\begin{equation}}
\def\ee{\end{equation}}

\def\L{{\cal L}}

\def\Lc{\mathscr{L}}

\def\a{\mathfrak{a}}
\def\b{\mathfrak{b}}

\def\d{\mathfrak{d}}
\def\e{\mathfrak{e}}
\def\f{\mathfrak{f}}

\begin{document}
\allowdisplaybreaks[1]
\renewcommand{\thefootnote}{\fnsymbol{footnote}}

\begin{titlepage}
\begin{flushright}
     CERN-2010  LPTHE-2010
       \\
\end{flushright}
\begin{center}
{{\Large \bf
 Four twisted differential operators \\for the  N=4 superconformal algebra
   }}
\lineskip.75em
\vskip 3em
\normalsize
\lineskip.75em \vskip 3em \normalsize {\large Laurent Baulieu
$^{  }
$\footnote{email address: baulieu@lpthe.jussieu.fr}
\\
\vskip 1em
{\it Theoretical Division CERN}\footnote{
CH-1211 Gen\`eve, 23, Switzerland }
\\
{\it LPTHE  Universit\'e Pierre et Marie Curie
}\footnote{ 4 place Jussieu, F-75252 Paris
Cedex 05, France}
\\
 }

\vskip 1 em
\end{center}
\vskip 1 em
\vskip 1 em
\begin{abstract}

\end{abstract}
The  $\mathcal{N}=4, d=4$ Yang--Mills conformal supersymmetry exhibits a very simple sub-sector described by  four differential operators. 
 The invariance under  this    subalgebra is big enough to  determine the $\mathcal{N}=4$ theory. Some attempts are done to interpret these differential operators.
\end{titlepage}
\renewcommand{\thefootnote}{\arabic{footnote}}
\setcounter{footnote}{0}



\renewcommand{\thefootnote}{\arabic{footnote}}
\setcounter{footnote}{0}



\def\N{N}

\def\a{\alpha}
  \def\d{\delta}
  \def\e{\eta}
  \def\x{\kappa}

\def\ni{\noindent}
\def\nn{\nonumber}
\def\be{\begin{eqnarray}}
\def\ee{\end{eqnarray}}
\def\e{\epsilon}
\newtheorem{theorem}{Theorem}
\newtheorem{corollary}[theorem]{Corollary}
\newpage
\section{Introduction}

The N=4 super-Yang--Mills action
\bea\label{n=4action}
  S=\frac{1}{g^2}\int d^4 x tr\big( \frac{1}{2}F_{\mu\nu}F^{\mu\nu}-i\bar{\lambda}^{\dot{\alpha}}_A\displaystyle{\not}D_{\dot{\alpha}\beta}\lambda^{\beta A}-i\lambda^{A}_{\alpha}\displaystyle{\not}D^{\alpha\dot{\beta}}\bar{\lambda}_{ A\dot{\beta}}+\frac{1}{2}(D_{\mu}\bar{\phi}_{AB})(D^{\mu}\phi^{AB})   \CR
 -\sqrt{2}\bar{\phi}_{AB}\{\lambda^{\alpha A},\lambda^{B}_{\alpha}\}-\sqrt{2}\phi^{AB}\{\bar{\lambda}^{\dot{\alpha}}_A,\bar{\lambda}_{\dot{\alpha}B}\}+\frac{1}{8}[\phi^{AB},\phi^{CD}][\bar{\phi}_{AB},\bar{\phi}_{CD}]\big)
\eea
is an interacting theory between a Yang--Mills field $A_\mu$, 6 scalar fields $\phi^{AB}$ and    4 Majorana spinors $\lambda^A = (\lambda ^{A\alpha}, \lambda^A_{\dot{\alpha}} )     $, where the indices $A,B$ label the $R$-symmetry. It  is invariant  under the Poincar\'e supersymmetry transformations with 16 supersymmetry transformations,
$\delta_\epsilon A^\mu =\bar{\epsilon} \gamma^\mu \lambda,\;\delta_\epsilon \phi^{AB}=\epsilon^A\lambda^B $ and 
$\delta  \lambda  = ( \Gamma^{\mu\nu} F_{\mu\nu}
+i D_\mu \Gamma^\mu\phi - [\phi,\phi])
  \epsilon  $. 
The N=4 theory is obtained by dimensionally reducing  the N=1 d=10 theory or, equivalently, the N=2 d=8 theory.
The 16 supersymmetries as a whole close only up to field equations, as follows,
\bea\{\delta_\epsilon, {\delta}_{\hat{\epsilon}}\}\sim-2i\bar{\epsilon}\gamma^\mu\hat{\epsilon}\partial_\mu-2i\delta^{gauge}(\bar{\epsilon}\gamma^\mu A_\mu\hat{\epsilon})\eea
where the symbol $\sim$ means modulo (spinor) field equations of motion.
 In a  light-cone   approach, one can  select 
  eight supersymmetries that close off-shell. However,  in this physical approach, Lorentz invariance is   difficult to recover      and, moreover,  light-cone gauge field  propagators are ill-defined, even in  perturbative quantum field theory.  In Euclidean space,  the Lorentz symmetry is  $SO(4)$ and   the   $R$-symmetry of the theory is $SO(5,1)$ (instead of $SO(6)$ in Minkowski space).  It is in fact possible to  reduce the global $SO(4)\times SO(5,1) $ invariance down to a $SO'(4)\times SL(2,R) $  invariance,  a so-called twist operation~\cite{Vafa}\cite{marcus}. In this way the fermion  and scalar representations  becomes reducible, and 
   one can     covariantly select   nine supersymmetry generators \cite{N=4} that close ``off-shell",  that is, constitute  a reduced super-algebra where no field equations occur in the closure relations  of transformations.   
Moreover,    the gauge transformation in the right-hand side of the closure relations can be eliminated, provided one introduces new fields called shadow fields~\cite{shadow}.  Eventually, one can do a gauge-fixing of the theory, such that one has Ward identities that allow one to control both gauge invariance and supersymmetry at the quantum level. This provides a solid framework for studying the supersymmetric properties of the N=4 theory within the framework of  quantum field theory~\cite{shadow}.

 The twist \cite{Vafa}\cite{marcus}  of the  N=4 superPoincar\'e symmetry   is done by taking the diagonal  $ SU'(2)$  of one of the  $ SU(2)\subset SO(5, 1)$  subgroups of the  $R$-symmetry  (there are 3 possible choices \cite{marcus})   and one of both    SU(2) factors   of the Lorentz group. What is left is  the  new Lorentz group $SO'(4)=SU(2)_L \otimes SU'    (2)$  and a part  of the  $R$-symmetry that contains at least a $U(1)$ symmetry, for instance,      a
    $SL(2,R)$  symmetry in the first twist.  Under the new Lorentz group, the supersymmetry generators  become scalars, vectors and (anti)self-dual tensors. The 16 super-charges are decomposed as follows,
\bea    (Q^{A}_{\alpha},\bar{Q}_{A\dot{\alpha}})\ \  \rightarrow \ \  (  Q_0,\bar{Q}_0, Q_\mu, {\bar Q}_\mu,{ Q_{\mu\nu^-}},\bar{Q}_{\mu\nu^-})\eea
 The nine charges  $Q_0,\bar{Q}_0, Q_\mu$ and $   Q_{\mu\nu^-}$ turn out to build an off-shell closed algebra, and, moreover, they can be geometrically constructed  from the point of view of topological field theory  \cite{N=4} as anticipated in~\cite{Baulieu.1998}.
Furthermore, the $N=4$ action is uniquely defined by the invariance under the symmetry with  the 6 generators $Q_0,\bar{Q}_0 $ and    $ Q_\mu$.  The 10=16-6 other supersymmetry generators are  overdetermining and appear as     accidental, but welcome,  symmetries of the N=4 action. For proving the finiteness of 1/2 BPS operators, one only need the 5-generator subalgebra made of $Q_0,\bar{Q}_0 $ and $   Q_{\mu\nu^-}$\cite{shadow}. 
This  may  challenge us to  find direct ways of constructing these maybe  more fundamental   smaller symmetries of the N=4 theory.  
The aim of this paper is to show that   a non-trivial part of the superconformal symmetry can be  also  directly built by generalizing  the framework of reference \cite{N=4}.
 This will provide    a  much smaller number of generators than
 the 32 ones  of  the twisted superconformal algebra \cite{hull}. It     determines the N=4 action, while keeping track of a non-trivial part of the superconformal algebra.  

  \section{   The superPoincar\'e twist of the N=4 Yang--Mills theory  }
  
To proceed, we need to give more details for the superPoincar\'e structure.  In the   first  twisted formalism, the gluino decomposes analogously  as the supersymmetry generators\footnote{It  is  convenient to identify  antiselfdual 2-forms      as  $SU(2)\subset SO'(4)$ valued scalars,       using the three independent   Kahler
 2-forms $J^I_{\mu\nu^-}$, ($I=1,2,3$), according to the invertible identity
$  X^I    \equiv J^I_{\mu\nu^-}X^{\mu\nu^-} $, so that
$h_{\mu\nu^-}\sim h^I$.},
    \begin{eqnarray}\label{twistedgluini}
 (\lambda^{a}_{\alpha}, {\lambda}_{a\dot{\alpha}})\ \ \ \ \rightarrow \ \ \ \  (\eta,\bar{\eta}, \psi_\mu,\bar{\psi}_\mu,\chi_{\mu\nu^-},\bar{\chi}_{\mu\nu^-})
 \equiv(\eta^\alpha, \psi^\alpha, \chi^{I\alpha})
\end{eqnarray}
where the $SL(2,R) $ indices  $1\leq \alpha\leq 2$ label the barred and unbarred fields.  So, after the twist,  the   $16=4\times 4$  spinorial  degrees of freedom of the conventional theory   are expressed as  $16 =(2+2\times 4+2\times3)$  $SO'(4)$-tensor degrees of freedom.

The   6 components of the  $SO(5,1)$-valued     scalar are twisted as follows
(the indices $i$ label the SL(2,R) adjoint representation),
\bea \phi  \ \rightarrow  (\ \phi^i, h_{\mu\nu^-})\equiv (\phi^i,h^I)\eea


 The symmetry  with  the nine generators $Q_0,\bar{Q}_0, Q_\mu,   Q_{\mu\nu^-}$    closes off-shell, by including among the fields   an auxiliary fields with 7 components, organized as  a vector $T_\mu$  and  a selfdual  2-form     $H_{\mu\nu^-}$.
 The   balanced system of fields,  denoted as    (9,16,7) multiplet,  is
\bea
(
 A_\mu,  \phi^i, \ h _{ \mu\nu^-}, \Psi   _\mu,  \bar  \Psi   _\mu,    \chi _{ \mu\nu^-},        \bar     \chi _{ \mu\nu^-},   \eta, \ \bar  \eta,      \
   T_\mu,\ H _{ \mu\nu^-})
\eea
A brute force change of variables  of  the  known  on-shell $N=4$ transformation can compute 
 the action on the fields of the   nine generators, with on-shell closure. Standard physicist methods can provide    their modifications to get off-shell closure by introducing     the auxiliary fields $T_\mu,H_{\mu\nu^-}$. 
There is in fact  a direct, and maybe more profound, construction  that  we will explain, since it   suggests to us the way to incorporate superconformal transformations and determine straightforwardly an interesting  off-shell closed sub-sector of the superconformal algebra.

   To find the  relevant supersymmetries, it  is best to start from eight dimensions,   using  the  TQFT  methods, and to   compactify the results in four dimensions. Indeed,  in  eight dimensions,  triality  indicates  immediately    the possibility of mapping
the 16  supersymmetry spinorial  generators  on twisted tensor generators,   as follows, \bea   (Q^\alpha, Q_{\dot\alpha})  \to ({ Q_0,  Q_M}, Q_{MN^-})
\eea
where   $1\leq M,N \leq 8$ are 
$SO(8)$ indices and  the   self-duality index ${MN^-}$  is  defined by     using the    $Spin(7) \subset SO(8)$-invariant selfdual  tensor
$t_{MNPQ}$. One has     $t_{8abc}= c_{abc}  $ where  the  $c_{abc}$'s are the    octonion structure coefficient.  Using this 4-tensor,      any given  $SO(8)$ 2-form    can be decomposed  as
$28=7\oplus 21$, in a $Spin(7) \subset SO(8)$-invariant way.   Thus $Q_{MN^-}$ stands for 7  generators.

The   9=1+8  generators $Q_0$ and $Q_M$ can be    determined explicitly from  the methods of TQFT~\cite{Baulieu.1998}\cite {BBT}. They satisfy
\bea Q_0^2=Q_M Q_N+ Q_N Q_M=0, \ \ \
 Q_0 Q_M+ Q_N Q_0 =\partial_M \eea
 This equation  implies off-shell closure, modulo gauge transformations and is obtained thank's to the introduction of an  auxiliary field that   is   a self-dual 2-form $T_{MN^-}$.   In seven dimensions, it   becomes   a    7-vector auxiliary field $T_a$,
$1\leq a\leq 7$.
 The non-closure relations are cornered in  the sector of the selfdual generator $Q_{MN^-}$.  Getting rid of $Q_{MN^-}$, one has   9=1+8 off-shell closed generators, which  is a  property that   survives  after dimensional reduction, e.g., for the N=4, d=4 theory.

The  off-shell closed representation of the $N=2, d=8$   theory is thus given by the balanced (9,16,7) multiplet
\bea\label{eightm}
( A_M, \Phi, \bar \Phi,\Psi   _M,    \chi _{ MN^-},     \eta,  T _{ MN^-})
\eea
This  8-dimensional formulation exists in curved space, provided the manifold has $Spin(7)$ holonomy, that is, one has   a constant spinor, which allows one to map all spinors on forms \footnote{ In fact the existence of such a constant spinor  $\zeta$ warrantees the existence of the  $Spin$(7)$\subset Spin(8)$-invariant  tensor
$t_{MNPQ}   =^t \zeta\Gamma_{MNPQ}  \zeta\   $, which allows one to split any given 2-form in  a selfdual and antiselfdual 2-form.  }. This is the    triality property.  In flat space, it can be understood  as a mere    $Spin$(7)$\subset Spin(8)$ invariant  changes of variables, using the  invariant tensor $t_{MNPQ}$. So, the 8-dimensional twist, we are concerned with, only preserves the $Spin$(7)$\subset Spin$(8) invariance.

One can then dimensionally reduce all formula in seven dimensions, with
 $A_M\to (A_a, L)$, $\Psi_M\to (\Psi_a, \bar\eta)$,
$\chi _{ MN^-}\to\bar\Psi _a$, $T_{ MN^-}\to\bar T _a$, where $a=1,\ldots,7$ is a $Spin(7)$ vector index. In seven dimensions,   the balanced 8-dimensional    multiplet (\ref{eightm}) becomes  the following one
\bea  (
 A_a, \Phi^i,  \Psi ^\alpha  _a,       \eta^\alpha,  T _{a}  )
\eea
The   $SL(2,R)$ indices $\alpha=1,2$ and $i=1,2,3$ arise naturally in the dimensional reduction, giving  a $Spin(7)\times SL(2,R)$ covariance  \cite{N=4}.
The   8 generators $Q_M$ become  a 7-vector   $Q_a$ and a scalar $\bar Q_0$.  The 7 generators 
$Q_{MN^-}$ become  another 7-vector  $\bar Q_a$, which enforces a global  $SL(2,R)$ covariance of the algebra, by pairing together
 $  Q_a$ and $\bar Q_a$.
 The non-closure relations are now cornered in the anti-commutation relations  between the 7-vector generators   $\bar Q_a$   and   $  Q_a$,   by
  equations of motion  that  appear in
  $\{\bar Q_a,   Q_b\}$,   proportionally  to the  antisymmetric octonionic tensor  $c_{abc}$.  By  introducing a seven-dimensional vector parameter $k^a$,  which is shared by both  vector generators $\bar Q_a$   and   $  Q_a$, 
 and  two  independent scalar parameters
$k^0$ and $\bar k^0$ for  both $\bar Q_0$   and   $  Q_0$,
one finds    an  off-shell closed    algebraic structure   for the four differential operators
$k^0Q_0,\bar k^0\bar Q_0, k^a Q_a $ and    $k^a\bar Q_a,$.  Their action on the fields  expresses an off-shell  closed supersymmetry   in seven dimensions, with 9=1+1+7 independent parameters.

One can   do a further dimensional reduction in four dimensions. The 7 auxiliary field $T_a$ decompose  into a vector and a self-dual tensor in four dimensions $(T_\mu,   H_{\mu\nu^-})$, and the 7  generators $Q_a$ decompose  into  a vector and a self-dual tensor. The nine off-shell closed generators are $Q_0,\bar{Q}_0, Q_\mu,Q_{\mu\nu^-}$.

One can in fact do a more subtle selection of generators, to reestablish the $SL(2,R)$  covariance in four dimensions. One    decomposes the 
 seven  generators ${\bar Q_a}$ into a vector and a self-dual tensor in four dimensions, and one retains the  $SL(2,R)$  covariant set of  10 generators, $Q_0,\bar{Q}_0,   Q_\mu$ and $  \bar Q_{\mu }$. The 6 generators $Q_0,\bar{Q}_0$ and $  Q_\mu$ build an off-shell closed subalgebra, but  the   off-shell closure is broken   between
 $Q_\mu$ and $ \bar Q_{\mu }$, by equations of motion proportional to  the antisymmetric tensor $\epsilon _{\mu\nu\rho\sigma}$. However, analogously  as in seven dimensions, one  can   introduce  a constant four-dimensional  vector parameter $k^\mu$ and  two scalar ones
$k^0$ and $\bar k^0$ and define
the 4 differential operators
$s^\alpha = (k^0 Q_0,\bar k^0\bar Q_0)$ and $\delta ^\alpha=(  k^\mu Q_\mu,k^\mu\bar Q_\mu)$,  $\a=1,2$, for a total  of  6=1+1+4 independent parameters.
One   can then set   $k^0=\bar k^0=1$ and one finds   the  following   $SL(2,R)$ and Lorentz covariant  graded  differential algebra in four dimensions\footnote{The notation $\delta_{gauge}(i_k A)$ stands for a gauge transformation with the field dependent  parameter $i_kA\equiv k^\mu A_\mu$, and the coefficients $\sigma _i^{\a\b} $ are for  a basis of   three $2\times2$  $SL(2,R)$ matrices. One also defines the graded Lie derivative   $\L_{\x} =i_\x d +di_\xi$. },
\bea\label{compact}
\{s^\alpha,s^\beta\}=\sigma^{i\alpha\beta}\delta_{gauge}(\Phi_i)\;,\; \{\delta^\alpha,\delta^\beta\}=\sigma^{i\alpha\beta}\delta_{gauge}(|k|^2\Phi_i)\;,\;\{s^\alpha,\delta^\beta\}=\epsilon^{\alpha\beta}(\mathcal{L}_k+\delta_{gauge}(i_kA))\CR\eea
This expresses in a   very compact    way  the off-shell closure   of maximal supersymmetry in four dimensions, with  equivariant closure relations.

A possible direct  and  geometrical construction of these relations  in the TQFT language  \cite{N=4} follows from  identities such as  
\bea (s+\delta+\bar{s}+\bar{\delta})(A+c) +(A+c)^2=\CR
 F+\psi+\bar{\psi}+g(k)(\eta+\bar{\eta})+g(J^I k)(\chi^I+\bar{\chi^I})+(1+|k|^2)(\bar{\Phi}+L+\Phi)
\eea

In view of these simplifications, one  can   consider a reverse construction. On can start from these anticommutation relations, solve them and  determine    the    transformation laws under $s^\alpha$ and $\delta^\alpha $ for the fields  $
 A\equiv A_\mu dx^\mu, \ \Psi ^\alpha\equiv\Psi _\mu^\alpha      dx^\mu, \ \chi^{I }_{ \alpha},\ \eta_\alpha, \ \Phi _i, \ h^I, \
   T\equiv T_\mu dx^\mu,\  H^I
   $.  
   
   For this, one uses power counting and grading conservation, as well as the covariance under the $SL(2,R)\times SO'(4)$ symmetry. By   denoting $g(k)\equiv g_{\mu\nu}k^\mu dx^\nu$ and $g(J^Ik) \equiv J^I_{\mu\nu}k^\mu dx^\nu$,  the solution of the algebra acting on the  fields of the balanced (9,16,7) 4-dimensional multiplet is
 \bea\nonumber
 \begin{split}
s^\alpha A &= \Psi^\alpha \\*
s^\alpha \Psi_\beta &= \delta^\alpha_\beta T -
{{\sigma^i}_\beta}^\alpha d_A \Phi_i \\*
s^\alpha \Phi_i &= \frac{1}{2} {\sigma_i}^{\alpha\beta} \eta_\beta \\*
s^\alpha \eta_\beta &= - 2 {{\sigma^{ij}}_\beta}^\alpha [\Phi_i,
\Phi_j] \\*
s^\alpha T &= \frac{1}{2} d_A \eta^\alpha + \sigma^{i\, \alpha \beta}[\Phi_i,
\Psi_\beta] \end{split} \hspace{10mm} \begin{split}
s^\alpha h^I &= \chi^{\alpha\, I} \\*
s^\alpha \chi^I_\beta &= \delta^\alpha_\beta H^I +
{{\sigma^i}_\beta}^\alpha [\Phi_i, h^I] \\*
s^\alpha H^I &= \frac{1}{2} [\eta^\alpha, h^I] + \sigma^{i\, \alpha\beta}
[\Phi_i, \chi^I_\beta]\end{split}
\eea
 \def\D{{d_A}}
\def\s{\sigma}
\bea \nonumber
\delta^\alpha A &=& g({\x}) \eta^\alpha  + g(J_I {\x})
\chi^{\alpha\, I} \CR
\delta^\alpha \Psi_\beta &=& \delta^\alpha_\beta \scal{i_{\x} F -
  g(J_I {\x}) H^I} + {{\sigma^i}_\beta}^\alpha g(J_I {\x})
[\Phi_i, h^I] - 2 {{\sigma^{ij}}_\beta}^\alpha g({\x}) [\Phi_i,
\Phi_j] \CR
\delta^\alpha \Phi_i &=& -\frac{1}{2} {\sigma_i}^{\alpha\beta}
i_{\x} \Psi_\beta \CR
\delta^\alpha \eta_\beta &=& -  \delta^\alpha_\beta
i_{\x} T +  {{\sigma^i}_\beta}^\alpha i_\x \D\Phi_i
\CR
\delta^\alpha T &=& \frac{1}{2} d_A i_{\x}
\Psi^\alpha - g(J_I {\x}) \scal{  [\eta^\alpha, h^I] + \sigma^{i\, \alpha\beta}  [\Phi_i, \chi^I_\beta]} + g({\x}) \sigma^{i\, \alpha\beta}
[\Phi_i, \eta_\beta] - \Lc_{\x} \Psi^\alpha
 \CR
\delta^\alpha h^I &=& -  i_{J^I {\x}} \Psi^\alpha \CR
\delta^\alpha \chi^I_\beta &=& \delta^\alpha_\beta \scal{
i_\x\D
  h^I
  + i_{J^I {\x}} T} +
{{\sigma^i}_\beta}^\alpha
i_{J^I\x}\D
 \Phi_i \CR
\delta^\alpha H^I &=&  \frac{1}{2} [ i_{\x} \Psi^\alpha,h^I] +
i_{J^I\x}\D
 \eta^\alpha +  \sigma^{i\, \alpha\beta}
[\Phi_i, i_
{J^I {\x}} \Psi_\beta] -
i_\x\D \chi^{\alpha\, I}
\eea
 One can then  show that the N=4 action~(\ref{n=4action})  is uniquely determined in twisted form by the  $s, \bar s $, $\delta$ (or $\bar\delta$ )  invariances, that is from a  symmetry with   6 parameters ($k_0,\bar k_0,k^\mu$).    Moreover, it  can be written as a $s\delta$-exact term \cite{N=4},
\begin{eqnarray}
I_{N=4}=\int\frac{1}{|k|}s\delta\left[g(k)(AdA+\frac{2}{3}A^3)+g(J^Ik)^*\epsilon_{IJK}h^Jdh^K+\bar{s}\bar{\delta}(\frac{1}{2}h^Ih_I-\frac{2}{3}\Phi_i \Phi^i)\right]
\end{eqnarray}
This action is independent on choice of  the constant vector  $\kappa$. An even more symmetrical expression of the action is
 \be \label{action} S_{N=4} =
- \frac{1}{2} \int_M \trace F_{\, \wedge} F + s^\alpha \delta_{ \alpha}
\mathscr{G}\ee
where
\be
 \mathscr{G}
= \int_M \trace \scal{ - \frac{1}{2} g({\x})_{\,
  \wedge} \scal{AF    - \frac{1}{3} A^3  } - \frac{1}{2} \star \varepsilon_{IJK}  h^I
i_{J^J {\x}} \D
h^K + \star \, s^\alpha \delta_{ \alpha} \, \, \scal{ \,
  \frac{1}{2} \, h_I  h^I\,  - \, \frac{2}{3} \, \Phi^i \Phi_i\,  }
 }\CR
\ee

 \section{Third twist and supersymmetric observables}
 The  passage by twist from the superPoincar\'e representation to the   first  twisted representation is  a linear mapping between fields, using 
 Pauli  matrices, giving equations that are invariant under a subgroup  $SO'(4) \times SL(2,R) \subset
 SO(4)\times SO(5,1)$  \cite{marcus}.
The third  twisted representation  can be in fact  obtained from  the first one  by the following invertible  $\kappa$-dependent  field redefinitions
  \be
 {V}_\mu  \equiv \kappa^\nu (h_{\mu\nu^-}    +g_{\mu\nu} L) \ \ \ 
 {\tilde \Psi}_\mu  \equiv \kappa^\nu (\bar \chi_{\mu\nu^-}    +g_{\mu\nu} \bar \eta  )
\ \ \ 
 {\bar  \Psi}_\mu  \equiv \kappa^\nu ( \chi_{\mu\nu^+}    +g_{\mu\nu}  \tilde\eta  )
 \ee
The  vector parameter $\kappa$, which  is necessary for  doing all changes of variables, disappears, modulo a boundary term,  when one  changes  variables from the   first   twisted Lagrangian to the third  twisted one.  The third  twisted variables are the most    most appropriate  to show the existence of supersymmetric variables, such as the supersymmetric Wilson-loop, which have interesting     finiteness properties  ~\cite{Zarembo}\cite{KW}\cite{Pestun}. The third twist formulation   has analogy with a complexified  expression of the twisted N=2,d=4 TQFT, with $A\to A+iV$, as sketched below. It is useful to give details on this in view of the forthcoming analysis of the conformal supersymmetry.
The set of fields in the third twist is  
 \be  (
 A_\mu,\ \Phi_i,\ h^I \  \ \Psi^\a_\mu,\ \eta^\a,\  \chi^{I\a}, \ T_\mu,\  G^I    )\to (
 A_\mu,\  V_\mu, \  \Phi,\  \bar \Phi\ \   \ \  \Psi _\mu,  \  \tilde \Psi_\mu, \   \chi_{\mu\nu^\pm},    \  \eta, \ \tilde\eta\ \ \ ,    H_{\mu\nu^\pm},\ H  )
 \ee
It has only an internal $U(1)\times SO(4)$ covariance.
 One has a $Q$-symmetry that can be recognized as  a combination of two of the    symmetries $s$ and $ \delta$,   governed by two parameters $u$ and $v$. It  can be shown to satisfy the complex equation \cite {complexN=2}
 \bea\label{horcomplex}
   (Q+d)  (A+iV+c )  +(A+iV +c)  = F_{A+iV}  +(u-iv)(\Psi +i\tilde \Psi)  +(u^2+v^2) \Phi
\eea
By defining
\be
Q\chi _\pm =  H_\pm -[c, \chi_\pm]
\ee
one finds that the  $N=4$ action  can be recomputed as a $Q$-exact term
\be
I=\int  d^4x  \frac{1}{u^2+v^2}   Q
 Re
[ \ \chi_{-} +i\chi_{-})
 ( u+iv) (F_{A+iV}   +    H_-+  i  H_+) +\ldots \ ]
\ee
The action is independent on $u$ and $v$. On can restrict to a particular $Q$-symmetry,  by  setting $u=iv$~\cite{KW}. For this restriction of the parameters, the  $Q$-transformations  for $A$ and $V$ are given by  
\be
Q(A+iV) =  -D_{A+iV}\ c= \delta_{gauge} (c) (A+iV)
\ee
Therefore the Wilson loop with argument $A+iV$ is  automatically $Q$-invariant
\be
Q exp \int dx^\mu (A_\mu+iV_\mu) =0
\ee
as well as any given gauge-invariant functional of $A+iV$. This  defines   $Q$-supersymmetric   observables for $u=iv$, which have been extensively studied in~\cite{KW}.
 These supersymmetric Wilson loops    can be    expressed in the first twist formulation, since the latter is related to the third twist by a mere  $\kappa$-dependent change of variables,  which leaves invariant the action.  It follows that   the mean value
\be
< exp \int_\Gamma  dx^\mu (A_\mu+i\kappa^\nu (J^J_{\mu\nu}    h^I     +g_{\mu\nu} L)) >
\ee
is independent on all possible local deformations, in particular on those of the contour $\Gamma$.

\section {Stochastic quantization and relation to Chern Simons action}
This section is devoted to a possible interpretation of the scalar generators of the N=4 theory as the   scalar supersymmetry of the stochastic quantization of a three-dimensional theory. 
The expression of the  $s\delta$-antecedent of     the N=4 action   (\ref{action})     strongly suggests   the influence of a three-dimensional  Chern simons action for the  four dimensional theory. In fact,  the general ideas of 
  stochastic quantization  \cite{stochastic} formally indicate  that,  if the contour is three-dimensional, one can either use the three-dimensional action
 \be\label{twist1}
 \int d^3x   (AdA+2/3 A^3  +    ^* \epsilon _{IJK} h^I  D ^J h^K  +^*L  D_I h^I  )
 \ee
 or the complex one 
 \be  \label{twist3}
 \int d^3x   ((A+iV)d(A+iV)+2/3 (A+iV)^3   )
 \ee
to compute certain 3-dimensional observables for the N=4 theory. The   action (\ref{twist1}) has only real gauge invariance while the action (\ref{twist3})    has complex gauge invariance  \cite{cswitten}, that is a double gauge symmetry.  In fact, the   former action  differs from the later one   by a covariant  gauge-fixing of the vector $V_\mu$.  We will formally show  that   stochastic quantization of  both    actions leads one to    supersymmetric theories, whose actions are  identical  either to    the   first  or to the   third  twisted N=4 actions, modulo $Q$-exact terms. The latter terms are irrelevant for the computation of $Q$-invariant observables, such as the above mentioned Wilson loops

In flat or curved Euclidian three-dimensional space one can define quantization by introducing a fourth (stochastic) time.  The time evolution is  then governed  by a Langevin equation.  Its  drift force  is the sum   of a force  along gauge-invariant directions,  which  is the equation of motion of the gauge-invariant action,    and of a force  along gauge orbits,  which  is equal to a gauge transformation where the parameter (real or complex) can be any given  arbitrary   function, possibly field dependent. The later  parameter  can be promoted to an independent  field over which one can functionally integrate.  The  reasoning is that   the expectation values  of  gauge-invariant  three-dimensional observables donnot depend on the choice of the parameters  of the drift forces along gauge orbits,  so that one  can consider a summation over these fields, since it yields  no modification of the value of gauge-invariant  observables. This is how gauge covariance can be enforced in the fourth dimension.   The additional  fields become the fourth component $A_0$  of the gauge field for the   action (\ref{twist1}),  or   the fourth components $A_0$
and   $V_0$ of  both fields $A_i$ and $V_i$  for the   action (\ref{twist3}).   The actions (\ref{twist1}) and (\ref{twist3}) are thus expected to generate the N=4 action  in their first and third twist formulations, with a segregation of the fourth component $x^0$ as a  somehow irrelevant variable. However one must consider observables at equal time $x^0$, and takes the limit $x^0\to \infty$. Let us see how this can happen.

Taking  the action (\ref{twist3}), the covariant Langevin equations that govern its quantization  are    \bea {F_{0,i}   -\epsilon _{ijk}  F^{jk}   - [V_0,V_i] +   \epsilon _{ijk} [V_j,V_K]  =
        b_i}\ \ \ \ \ \  {  D_{[0}  V_{i]}    =  \bar  b_i}\eea
One can  express the Langevin process as path  integral with a Gaussian   dependance in the noises $b$ and $\bar b$,  doing   a change of variable between     $b,\bar b$ and $A$ and $V$. This    necessitates the insertion of     Jacobians. The latter   can be  expressed as a path integral over fermions, which will be  interpreted as the fermions of the twisted theory. Since one   has zero modes in the gauge covariant  Langevin equations, their  gauge-fixing introduces fermionic auxiliary fields,   with the occurrence  a super-Jacobian, which yields    a functional representation with the commuting scalar fields  $\Phi$ and $\bar\Phi$.  It goes as follows. One
  uses  4-dimensional notations, so that the Langevin equations can be rewritten as   
\bea {F_{\mu\nu^+}    - V_{[\mu}V_{\nu]^+}     =
        b_{\mu\nu^+}}
\ \ \ \ \ \ \  {  D_{[\mu}  V_{\nu]^+}    =  \bar  b_{\mu\nu^+}}\eea
One defines a new covariant equation  for the stochastic evolution of $V_0$ as
\bea
  D^\mu  V_\mu   =\bar b
\eea
By doing  standard   steps of  inserting delta  functions and determinants  in a Gaussian  path    integral  representation as in \cite{stochastic}, one ends up with the  following action for describing the Langevin process
\bea\label{action2}
I_{GF}=
\int dtdx{ Tr}  s _{top}     ( \chi _{\mu\nu^+}    ( F_{\mu\nu^+}  - V_{[\mu}V_{\nu]^+}     +  D_{[\mu}  V_{\nu]^+  }    -{1\over2} \bar   b_{\mu\nu^+})
\CR
+\bar  \chi _{\mu\nu^+}    ( F_{\mu\nu^+ } - V_{[\mu}V_{\nu]^+}
  -D_{[\mu}  V_{\nu]^+        }    -{1\over2}  b_{\mu\nu^+})
  +\chi  ( D _\mu V_\mu  -{1\over2}\bar  b)
+\bar \Phi  D_\mu \Psi_\mu
+\bar c (  \partial  _\mu A_\mu  -{1\over2}  b)  )\CR
  \eea
  where
  \bea s _{top}A_\mu    &=&\Psi_\mu +D_\mu  c   \ \ \ \ \ \ \   \ \ \ \ \ \ \  \ \ \ \ \ \ \  s_{top}c  = \Phi  -cc   \cr s_{top}\Psi_\mu  &=&  D_\mu  \Phi -[c, \Psi_\mu] \ \ \ \ \ \ \   \ \ \ \ \ \ \  \ 
s_{top}\Phi =  -[c,\Phi  ] \cr
s_{top}\chi _{\mu\nu^\pm}   &=& b_{\mu\nu^+}\ \ \ \ \ \ \   \ \ \ \ \ \ \  \ \ \ \ \ \ \  s_{top}   b_{\mu\nu^\pm}  =  0\CR
 s_{top}\bar{\Phi} &=&\eta \ \ \ \ \ \ \  \ \ \ \ \ \ \  \ \ \ \ \ \ \  \ \ \ \ \ \ \ s_{top}\eta =0 \CR
  s_{top}\bar{c} &=&b  \ \ \ \ \ \ \    \ \ \ \ \ \ \  \ \ \ \ \ \ \ \ \ \ \ \ \ \  s_{top}b=0
\eea

One can identify  
$Q$ and $ s _{top}$ and the action~(\ref{action2}) is identical to the N=4 theory in the third twist, modulo $Q$-exact terms. The latter  terms contain  the quartic scalar field interactions. Their omission  does not  change the expectation values for $Q$-invariant observables.  One obtains a similar result for the first  twist,   starting  from  the three-dimensional action  (\ref{twist1}) for the fields $A_i, L, V_i$, which yields the  action of the N=4 theory in the first twist, modulo $Q$-exact terms.

     There is no obstruction to do this formal construction in curved three-dimensional space. Moreover,  one can introduce  non-flat metrics components $g_{oi}$, which could maybe ease  certain practical computations. We  leave open the problem of directly computing $Q$-invariant  Wilson loops of the four-dimensional  theory with three-dimensional  contours, directly in the Chern--Simons three-dimensional theory.

We see that the N=4 theory relies on building blocks that are much more elementary than expected. In what follows, we will see that one can extend these idea, and incorporate elements of special supersymmetry from the beginning. It is indeed interesting to introduce the superconformal algebra  in a constructive way, with no redundancy.

%
%
%
%
%
%

  \section{Inclusion of part of the conformal symmetry in the $Q$ symmetry}

  The  conformal Yang--Mills supersymmetry  is governed  by 32 generators, with spinor parameters $\epsilon $ and $\eta$,
  \begin{eqnarray} \nonumber
\delta A_\mu &=&  \lambda{\Gamma}_\mu (  \epsilon +x^\mu \gamma_\mu  \eta )\nn\\
\delta \vec \varphi &=& (  \epsilon +x^\mu \gamma_\mu\eta )\vec \tau
\lambda
 \nn\\
\delta  \lambda  &=&  ( \Gamma^{\mu\nu} F_{\mu\nu}
+i D_\mu \Gamma^\mu\varphi - [\varphi,\varphi])
 (  \epsilon +x^\mu \gamma_\mu  \eta )+2i   \varphi \eta\nonumber
\end{eqnarray}

 In \cite{hull}, this  superconformal  symmetry has been twisted by  reducing  the product of its $R$-symmetry and  conformal symmetry $SO(5,1)\times SO(5,1)$,      in a way that generalises the mixing between the Lorentz symmetry and the  $R$-symmetry for the superPoincar\'e case.  
 
We will follow a different root. We    generalize the algebra~(\ref{compact}) for  the four scalar generators, by replacing the constant vector $k^\mu$ into  a local one that is proportional to the coordinate $x^\mu$.  We retain  the     same field representations as in the first twist.
   Some compensating transformations must  be    done for absorbing non-homogeneous terms in $x^\mu$, using the existing global  symmetries.
  After some thoughts, one concludes that one must consider   the following distorted algebra   
\def\x{x}
 \def\a{\alpha}
  \def\d{\delta}
  \def\e{\eta}
  \def\w{w}
    \def\f{\frac{1}{2}}
     \def\f{{ a}}
   \be    \{ s^\alpha,  s^\beta \} = 2\sigma^{i\,
  \alpha\beta} \delta_{\mathrm{gauge}}(\Phi_i)
 \ \ \  \ \ \ \ \{ \delta_x^\alpha
,\delta_x^\beta \} = 2
  \sigma^{i\,\alpha\beta} \delta_{\mathrm{gauge}}(  |{\x} |^2  \Phi_i) \ee
 \be  
  \{ s^\alpha, \delta_{x}^\beta \} = \varepsilon^{\alpha\beta}
\scal{ \L_{\x} + \delta_{\mathrm{gauge}}(i_{\x} A)   + {   \w }  }
 +
{
  \Delta _{SL(2,R)}^{ {\alpha \beta} }}
\ee  
 $\Delta _{SL(2,R)}$ is a  ${SL(2,R)}$ transformation and
 $\w$ is a   $U(1)$ symmetry, which counts  the conformal weight of fields. The requirement of $SO(4)\times SL(2,R)$ covariance and the respect of the various gradings determine the structure of  this algebra. One must look  for possible representations, in terms of fields, and one  is led to check whether
some  constants ${  A, Z,W,W'}$  exist,  such that the following transformation laws fulfill    the above anti-commutation relations.
\bea
\label{BRST} \begin{split}
s^\alpha A &= \Psi^\alpha \\*
s^\alpha \Psi_\beta &= \delta^\alpha_\beta T -
{{\sigma^i}_\beta}^\alpha d_A \Phi_i \\*
s^\alpha \Phi_i &= \frac{1}{2} {\sigma_i}^{\alpha\beta} \eta_\beta \\*
s^\alpha \eta_\beta &= - 2 {{\sigma^{ij}}_\beta}^\alpha [\Phi_i,
\Phi_j] \\*
s^\alpha T &= \frac{1}{2} d_A \eta^\alpha + \sigma^{i\, \alpha \beta}[\Phi_i,
\Psi_\beta] \end{split} \hspace{10mm} \begin{split}
s^\alpha h^I &= \chi^{\alpha\, I} \\*
s^\alpha \chi^I_\beta &= \delta^\alpha_\beta H^I +
{{\sigma^i}_\beta}^\alpha [\Phi_i, h^I] \\*
s^\alpha H^I &= \frac{1}{2} [\eta^\alpha, h^I] + \sigma^{i\, \alpha\beta}
[\Phi_i, \chi^I_\beta]\end{split}
\eea
 %
\bea\label{sconf}
\delta_x^\alpha A &=& g({\x}) \eta^\alpha  + g(J_I {\x})
\chi^{\alpha\, I} \CR
\delta_x^\alpha \Psi_\beta &=& \delta^\alpha_\beta \scal{i_{\x} F -
  g(J_I {\x}) H^I} + {{\sigma^i}_\beta}^\alpha g(J_I {\x})
[\Phi_i, h^I] - 2 {{\sigma^{ij}}_\beta}^\alpha g({\x}) [\Phi_i,
\Phi_j] \CR
\delta_x^\alpha \Phi_i &=& -\frac{1}{2} {\sigma_i}^{\alpha\beta}
i_{\x} \Psi_\beta \CR
\delta_x^\alpha \eta_\beta &=& -  \delta^\alpha_\beta
i_{\x} T +  {{\sigma^i}_\beta}^\alpha i_\x \D\Phi_i   +
{  A   {\sigma_i}^{\alpha\beta}  \Phi^i  }
\CR
\delta_x^\alpha T &=& \frac{1}{2} d_A i_{\x}
\Psi^\alpha - g(J_I {\x}) \scal{  [\eta^\alpha, h^I] + \sigma^{i\, \alpha\beta}  [\Phi_i, \chi^I_\beta]} + g({\x}) \sigma^{i\, \alpha\beta}
[\Phi_i, \eta_\beta] - \Lc_{\x} \Psi^\alpha
{   +Z
  \delta^\alpha_\beta    \Psi  ^\beta  }
 \CR
\delta_x^\alpha h^I &=& -  i_{J^I {\x}} \Psi^\alpha \CR
\delta_x^\alpha \chi^I_\beta &=& \delta^\alpha_\beta \scal{
i_\x\D
  h^I { + W h^I}  + i_{J^I {\x}} T} +
{{\sigma^i}_\beta}^\alpha
i_{J^I\x}\D
 \Phi_i \CR
\delta_x^\alpha H^I &=&  \frac{1}{2} [ i_{\x} \Psi^\alpha,h^I] +
i_{J^I\x}\D
 \eta^\alpha +  \sigma^{i\, \alpha\beta}
[\Phi_i, i_
{J^I {\x}} \Psi_\beta] -
i_\x\D \chi^{\alpha\, I}
{ +
W' \chi  ^{I \a}}  
\eea

One gets after a lengthy computation that there is indeed a unique solution, given by
$Z=-1, A = 2, W=1, W' = 2$.

One   thus obtains the intriguing result that,     ``special",  i.e., x-dependent $\delta_x$ transformations, exist that are very simply related to the  twisted vector super Poincar\'e supersymmetry transformation,   as follows
   \bea
 \delta_x =x^\mu  (\delta _\mu + g_{\mu\nu} \frac {x ^\nu}{x^2} C )
 \eea
Here the operator  $\delta _\mu $   is identical to that of the vector supersymmetry of the superPoincar\'e algebra, and
 the operator  $C$ is the further modification brought by the special supersymmetry, as it is implied by the graded commutation  relations. The action of $C$  is only non-zero for $\eta^\a$,  $\chi^{I\a}$,  $T$  and $H^I$ and can be read from Eqs.~(\ref{sconf}).

One can verify that the above  4 symmetries can be identified as combinations of  the  twisted ones  that  are    obtained by computing the  first-twist of the 32 generators of the superconformal transformation \cite{hull}.  Here, they have arised in a somehow very elementary geometrical construction, and they capture an interesting part of the maximal  conformal supersymmetry with its 32 generators.
 Indeed, one can    verify that the N=4 action, in first twisted form is completely determined by its invariance  under  both graded Poincar\'e operators $s,\bar s$ and both  special  supersymmetry operators $\delta_x,\bar\delta_x$.

Does it help to discuss special supersymmetric observables as in \cite{Pestun}, and determine some of them?
 One can generalize  the trick that we used  by combining the  scalar and  vector symmetries for the ordinary supersymmetric observables, and define
 \be
 Q=u  s   +v\d_x
 \ee
 where the scalar parameters  $u, v $ are commuting ones.  One then finds that $Q$ satisfies an equation as in
 Eq.~(\ref{horcomplex}), by a simple comparison between the $\delta$- and  $\delta_x$- transformations. 

 It follows  that the following  1-form :
 \be
 A + \frac{1}{x^2}\scal { i_{J^J {\x}}  h^I +L} = dx^\mu \scal{ A_\mu + \frac{x^\nu}{x^2}\scal{  J^J_{\mu\nu}    h^I 
 +g_{\mu\nu} L} }
 \ee
    transforms under  $s+\bar\d $ simply by a gauge transformation when $u^\a= -iv^\a$, provided
  that 
  \bea
  {x^2}=1
  \eea 
  The verification is as in section 3, except that the constant  $\kappa^\mu$ has been replaced by $x^\mu$.
Therefore the following special Wilson loop is $s+\bar\d $ invariant :
 \be
  \scal{s+\bar\d  }   \exp i \int _{\Gamma_{  {x^2}=1}} \scal{  A + \frac{1}{x^2}\scal { i_{J^J {\x}}  h^I +L} } =0
 \ee
  Notice that because $  {x^2}=1$, the term $dx^\mu g_{\mu\nu} x^\nu L$ disappears on the contour. The Wilson loop invariance equation reduces  therefore to the following one      \be
  \scal{s+\bar\d  }   \exp i \int _{\Gamma_{  {x^2}=1}}  dx^\mu \scal{  A_\mu + \frac{1}{x^2}  { 
  h_{\mu\nu^-}x^\nu  } }  =0
 \ee
  So, the special supersymmetric Wilson loop must be defined   on a circle, and only depends on 3 of the scalar fields.
Its origin  is   analogous to that of the ordinary Wilson loop $
  \exp i \int _\Gamma \scal{  A + iV}
$  in the third twist. The reason is because the special supersymmetry has a lot in common with the twisted vector symmetry  and  because the manipulations  of using the third twist mapping  are   similar. One may 
     question  about  the  finiteness of observables,  
   the   topological properties, etc.. of such special observables that are invariant under $s+\bar \d$   and computed by mean of the $N=4$ action, which    we have shown is $s$ and $\bar \d$ invariant, and thus $s+\bar \d$-invariant.  However, the use of Ward identities is complicated by the lack of translational invariance, due to the dependance on x of the transformations, so the topological properties are presumably lost, and the observables  probably depend on the detail of the metrics. Interestingly, specific examples have shown that such observables are not automatically topological \cite{Pestun}.  Perhaps, computing these special Wilson loops within   a 3-dimensional framework  will be useful, as suggested by the results of stochastic quantization.

 \section*{Acknowledgments}
 I thank  
 G. Bossard and K. Zarembo  for    interesting and constructive   discussions on the subject.

\end{document}